\documentclass[a4paper,11pt]{article}
\usepackage{pos}

\newcommand{\bec}{\begin{center}}
\newcommand{\eec}{\end{center}}
\newcommand{\beq}{\begin{equation}}
\newcommand{\eeq}{\end{equation}}
\newcommand{\bea}{\begin{eqnarray}}
\newcommand{\eea}{\end{eqnarray}}

\newcommand{\hf}{\frac{1}{2}}

\newcommand{\qtr}{\frac{1}{4}}
\newcommand{\psib}{{\overline{\psi}}}

\title{Complex Langevin study of spontaneous symmetry breaking in IKKT matrix model}

\author*{Arpith Kumar}
\author{Anosh Joseph}
\author{Piyush Kumar}

\affiliation{Department of Physical Sciences, \\Indian Institute of Science Education and Research (IISER) Mohali, \\Knowledge City, Sector 81, SAS Nagar, Punjab 140306, India}

\emailAdd{arpithk.iiserm@gmail.com}
\emailAdd{anoshjoseph@iisermohali.ac.in}
\emailAdd{piyush.iiserm@gmail.com}

\abstract{The IKKT matrix model, in the large-$N$ limit, is conjectured to be a non-perturbative definition of the ten-dimensional type IIB superstring theory. In this work, we investigate the possibility of spontaneous breaking of the ten-dimensional rotational symmetry in the Euclidean IKKT model. Since the effective action, after integrating out the fermions, is inherently complex, we use the complex Langevin dynamics to study the model. In order to evade the singular-drift problem in the model, we add supersymmetry preserving deformations and then take the vanishing limit of the deformations. Our analysis suggests that the phase of the Pfaffian indeed induces the spontaneous SO(10) symmetry breaking in the Euclidean IKKT model.}

\FullConference{%
 The 39th International Symposium on Lattice Field Theory - LATTICE 2022\\
  8-13 August 2022\\
  Bonn, Germany
}

\begin{document}
\maketitle

\section{Introduction}
\label{sec:intro}
Non-perturbative studies of ten-dimensional superstring theories are essential to understand the emergence of spacetime. In particular, the dynamical compactification of six extra dimensions is critical for such theories to be phenomenologically admissible. Matrix models are standard tools to investigate the non-perturbative aspects of superstrings. The IKKT (type IIB) matrix model was proposed in 1996 as a constructive definition of the ten-dimensional type IIB superstring theory \cite{Ishibashi:1996xs}. The action is a matrix regularization of the type IIB superstring action in the Schild gauge \cite{Green:1983wt}. The zero-volume limit of the ten-dimensional $\mathcal{N} = 1$ super Yang-Mills with SU($N$) gauge group formally yields the IKKT matrix model. The equivalence between the IKKT matrix model and type IIB superstring holds in the large-$N$ limit. The ten-dimensional extended $\mathcal{N} = 2$ supersymmetry ensures that gravity is included. The $N \times N$ bosonic matrices are analogous to the gravitational degrees of freedom, where the eigenvalues of the matrices denote the spacetime points. In this model, spacetime does not exist a priori but is dynamically generated from the matrix degrees of freedom. In the large-$N$ limit, a smooth spacetime manifold is expected to emerge from the eigenvalues. The compactification of the extra dimensions suggests that the distribution of eigenvalues should collapse to a lower-dimensional manifold. When this occurs in the Euclidean signature, the ten-dimensional rotational symmetry of the model must be spontaneously broken. 

In this work, we investigate the possibility of spontaneous symmetry breaking (SSB) of SO$(10)$ symmetry in the Euclidean version of the IKKT matrix model. The model has a severe sign problem; the Pfaffian obtained after integrating out fermions is inherently complex. The phase of the Pfaffian plays a critical role in determining the correct vacuum of the model. Unfortunately, Monte Carlo methods are unreliable for studying complex action matrix models. In recent years, the complex Langevin method \cite{Klauder:1983nn, Parisi:1984cs} has emerged as a successful candidate for tackling models with the sign problem. While applying the complex Langevin method to the Euclidean IKKT matrix model, we encounter problems that hamper the reliability of the simulations. The singular-drift problem is one of them. To avoid this problem, we suggest introducing a supersymmetry-preserving deformation to the IKKT model. The original IKKT matrix model is recovered in the vanishing limit of the deformation parameter.

In this proceedings, we present our preliminary results from the complex Langevin analysis of the IKKT matrix model with Euclidean signature. In Sec. \ref{sec:ikkt-model}, we briefly discuss the mathematical formalism of the model and the associated sign problem. Sec. \ref{sec:clm-ikkt} explains the problems associated with the complex Langevin study of the model. We introduce supersymmetry-preserving deformations in Sec. \ref{sec:susy-def} and discuss the simulation results. Sec. \ref{sec:conclusion} is devoted to conclusions and future directions.

\section{Review of the Euclidean IKKT matrix model}
\label{sec:ikkt-model}

The Euclidean IKKT matrix model, obtained by a Wick rotation of the Lorentzian version, has a finite well-defined partition function \cite{Krauth:1998xh, Austing:2001pk},
\beq
\label{eqn:ikkt-action}
Z = \int dX d\psi e^{-S_{\rm IKKT}},
\eeq
where
\beq
S_{\rm IKKT} = S_\text{b} + S_\text{f}, ~~ {\rm with}~~S_\text{b} = -\qtr N ~\text{tr} \left([X_\mu, X_\nu]^2 \right)~~{\rm and}~~S_\text{f} = -\hf N ~\text{tr} \left(\psi_\alpha (\mathcal{C} \Gamma^\mu)_{\alpha\beta}[X_\mu,\psi_\beta] \right).
\eeq
The $N \times N$ traceless Hermitian matrices, $X_\mu (\mu = 1, 2, 3, \cdots,~ 10)$ and $\psi_\alpha (\alpha = 1, 2, 3, \cdots,~ 16)$ transform respectively as vectors and Majorana-Weyl spinors under SO(10) transformations. We consider the Weyl projected representation of gamma matrices $\Gamma_{\mu}$ in ten dimensions. In this representation, the charge conjugation matrix $\mathcal{C}$, satisfying $\mathcal{C} \Gamma^{\mu} \mathcal{C}^{\dagger} = \left(\Gamma^{\mu}\right)^{T}$ and $\mathcal{C}^{T} = \mathcal{C}$,  becomes an identity matrix. The action manifests SU($N$) gauge symmetry, extended $\mathcal{N} = 2$ supersymmetry, and SO(10) rotational symmetry. 
	
The partition function, after integrating out the fermions reads,
\beq
Z = \int dX ~ {\rm Pf} \mathcal{M}~ e^{-S_{\rm b}}= \int dX ~ e^{-S_{\rm eff}}; ~~ S_{\rm eff} = S_\text{b} - \ln  {\rm Pf} \mathcal{M},  
\eeq
where the fermionic operator, $\mathcal{M}$ is a $16(N^2 -1) \times 16(N^2 -1)$ anti-symmetric matrix. In order to get the explicit form of $\mathcal{M}$, we expand $X_\mu$ and $\psi_\alpha$ in terms of the $N^2-1$ generators, $\{t^a\}$ of ${\rm SU}(N)$  as follows
\beq
X_\mu = \sum_{a=1}^{N^2-1} X_\mu^a t^a ~~{\rm and}~~ \psi_\alpha = \sum_{b=1}^{N^2-1} \psi_\alpha^b t^b,
\eeq
where $X_\mu^a$ and $\psi_\alpha^b$ are real and Grassmann numbers, respectively. The traceless, Hermitian generators are normalized as ${\rm tr} \left( t^a t^b \right) = \delta^{ab}$.
Using the properties of SU($N$) structure constants, we have
\bea
\mathcal{M}_{\alpha a, \beta b} &=&  \frac{N}{2} \Gamma_{\alpha \beta}^\mu ~{\rm tr} \Big( X_\mu  \left[ t^a,t^b \right] \Big).
\eea
The interpretation of the eigenvalues as the spacetime points allows us to define the `radial extent’ of spacetime in each direction as follows
\bea
\langle \lambda_\mu \rangle = \Big \langle  \frac{1}{N} {\rm tr} \left( X_\mu^2 \right) \Big  \rangle.
\eea
We consider $\lambda_\mu$ as an order parameter for investigating SSB of SO(10) symmetry. In the large-$N$ limit, if these extents are not equivalent, i.e., if they grow along some dimensions, $d < 10$, and shrink along others, we say that the SO(10) symmetry spontaneously breaks down to SO(d). The bosonic IKKT model was studied using Monte Carlo method and $1/D$ expansion, and no SSB was observed \cite{Hotta:1998en}. Later, phase-quenched Monte Carlo studies were performed, and again no SSB was evident \cite{Ambjorn:2000dx, Anagnostopoulos:2012ib}. These studies point to the fact that the complex phase of the Pfaffian plays a crucial role in SSB. The phase fluctuates wildly, suggesting that the sign problem is severe; hence, phase-quenched approximations are inexact. There exist only a few methods that are capable of incorporating the complex phase and tackling the associated sign problem. The complex Langevin method is one such promising approach.

\section{Applying complex Langevin to the IKKT model}
\label{sec:clm-ikkt}

This section discusses the application of the complex Langevin method to the Euclidean IKKT model. The update of bosonic matrices $X_\mu$ at Langevin time $\tau$ reads
\bea
\frac{d(X_{\mu})_{ij}}{d\tau} = -\frac{\partial S_{\rm eff}}{\partial(X_{\mu})_{ji}} + (\eta_{\mu})_{ij}(\tau),
\eea
where $\eta_{\mu}(\tau)$ is a Hermitian Gaussian noise obeying the probability distribution $\exp \left(\qtr \int {\rm tr} \left( \eta^2_{\mu}(\tau) \right) \right)$. 

Sometimes, the complex Langevin method can give wrong results due to incorrect convergence. Fortunately, there exist certain correctness criteria \cite{Seiler:2012wz, Nagata:2016vkn}, which can validate the simulation results. The more recent one is based on the localized distribution of the probability of complex field configurations. The distribution of the magnitude of the drift 
\beq
u=\sqrt{\frac{1}{10N^2} \sum_{\mu=1}^{10} \sum_{i,j=1}^{N} \left| \frac{\partial S_{\rm eff}}{\partial(X_{\mu})_{ji}}\right|^2}
\eeq
should be suppressed exponentially or faster to ascertain the reliability of the simulations. 
\begin{figure}[htbp]
	\centering
	\includegraphics[width=.48\textwidth,origin=c,angle=0]{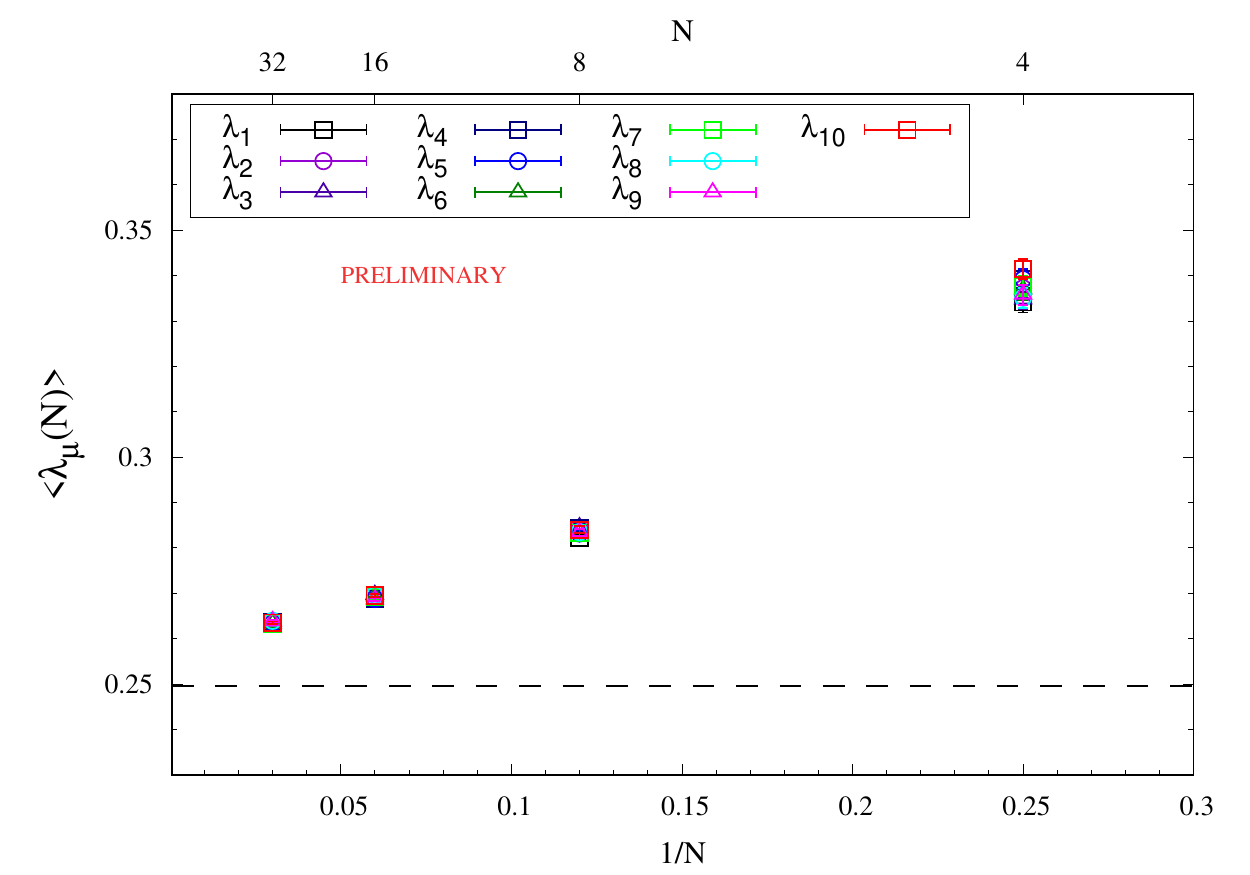}
	\includegraphics[width=.48\textwidth,origin=c,angle=0]{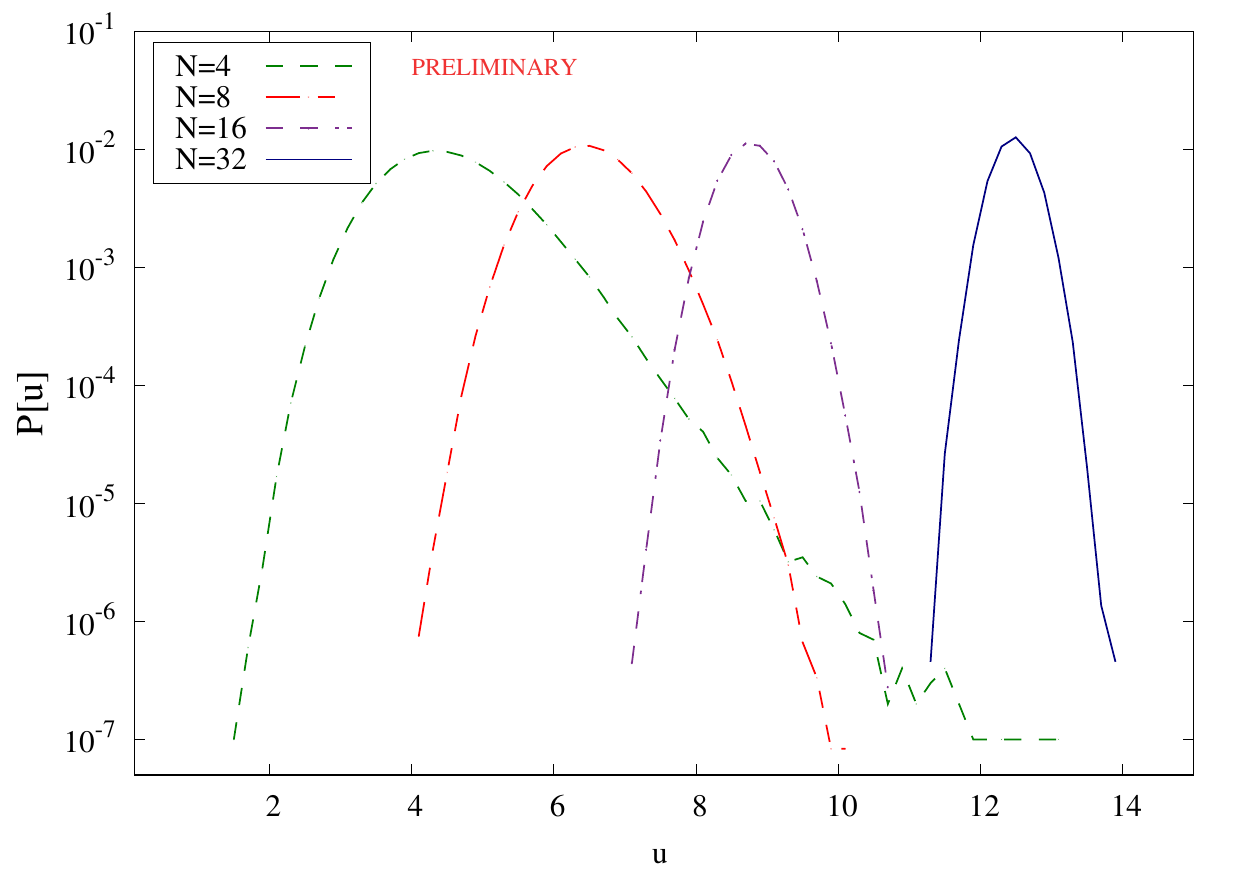}
	\caption{Bosonic IKKT model. (Left) The expectation value of order parameter $\lambda_{\mu}$ and (Right) the corresponding probability of the magnitude of the drift for various $N$.} 
	\label{fig:bos}
\end{figure}

In Fig. \ref{fig:bos} we present the complex Langevin simulation results for the bosonic IKKT model. From the plot on the left panel, we infer that the SO(10) symmetry is intact even for finite-$N$, and approaches the analytical value in the large-$N$ limit \cite{Hotta:1998en}. The probability of drift plotted on the right panel falls off exponentially or faster, which implies that the simulation results are reliable. While applying complex Langevin method to the Euclidean IKKT model, we encountered two major problems, namely the {\it excursion problem} and the {\it singular-drift problem}, that violate the above mentioned correctness criterion. In the upcoming subsections we briefly discuss these problems and the ways to circumvent them.

\subsection{Excursion problem and gauge cooling}
\label{subsec:excursion-problem}

The inherent complex nature of the Pfaffian can result in excursions of the bosonic matrices $X_\mu$ into anti-Hermitian directions, enlarging the group space from SU($N$) to ${\rm SL}(N, \mathbb{C})$. We encounter the excursion problem when $X_\mu$ wanders too far away from SU($N$). A proposed solution to this problem is gauge cooling \cite{Seiler:2012wz}. 
We define the `Hermiticity norm' \cite{Nagata:2016vkn}
\beq
\mathcal{N}_{\rm H} \equiv -\frac{1}{10N} \sum_{\mu}{\rm tr} \left( \left[ X_\mu - X_\mu^\dagger \right]^2 \right)
\eeq
to track the deviation of $X_\mu$ from Hermitian configurations. The matrix fields $X_\mu$ are invariant under the enlarged gauge symmetry,
\bea
X_\mu \rightarrow g X_\mu g^{-1},~ g \in {\rm SL}(N,\mathbb{C})  \\
g = {\rm e}^{-\alpha \delta{\mathcal{N}}_{\rm H}}, ~~ \delta{\mathcal{N}}_{\rm H} = \frac{1}{N} \sum_\mu\left[ X_\mu, X_\mu^\dagger \right], ~~\alpha \in \mathbb{R}^+,  
\eea
while $\mathcal{N}_{\rm H}$ is not invariant. We utilize this property and successively apply the above gauge transformation at each Langevin step until $\mathcal{N}_{\rm H}$ is minimized. The gauge cooling procedure has been proven to respect complex Langevin correctness criteria \cite{Nagata:2016vkn}. In our simulations, we observe that after applying gauge cooling, $\mathcal{N}_{\rm H}$ is well under control.  

\subsection{Singular drift problem and mass deformations}
\label{subsec:singular-drift-problem}

The gradient of the effective fermionic action contains the inverse of the fermion operator, $\mathcal{M}^{-1}$. The singular-drift problem arises when the eigenvalues of $\mathcal{M}$ accumulate densely near zero. One way to avoid this problem is to shift the eigenvalues of the fermion operator away from the origin. This shift can be introduced by adding fermion bilinear mass deformation terms to the action \cite{Ito:2016efb}. In general, the deformations have the following form
\bea
\Delta S = \frac{N}{2}\epsilon m_\mu  {\rm tr}\left( X_\mu ^2 \right) +  \frac{N}{2}  {\rm tr} \left( \psi_\alpha (\mathcal{CA})_{\alpha\beta} \psi_\beta \right),
\eea
where $m_\mu$ is the mass vector and $\mathcal{A}$ is a complex $16 \times 16$ anti-symmetric matrix. Majorana-Weyl spinors severely limit the allowed ranks of gamma matrices in ten dimensions. This implies that only bilinears of rank three and seven tensor (equivalent due to the duality relations) survive \cite{Wetterich:1982eh}, that is, $\mathcal{A} = i m_{\rm f} \epsilon_{\mu\nu\sigma} \Gamma_{\mu} \Gamma_{\nu}^\dagger \Gamma_\sigma$ with totally anti-symmetric $\epsilon_{\mu \nu \sigma}$ 3-form. Here $\epsilon$ and $m_{\rm f}$ are the deformation parameters. Apart from explicitly breaking the SO(10) symmetry, such deformations induce supersymmetry breaking. The extended $\mathcal{N} = 2$ supersymmetry is crucial for the model to include gravity. Similar deformations were considered in a recent study \cite{Anagnostopoulos:2020xai}, where the authors concluded that the SO(10) symmetry was spontaneously broken down to SO(3) (consistent with Gaussian expansion method results \cite{Nishimura:2011xy}). Studying SSB with this deformation requires three-step extrapolations, $N \rightarrow \infty$, $\epsilon \rightarrow 0$, $m_{\rm f} \rightarrow 0$, which introduce systematic errors. In this work, we suggest supersymmetry-preserving deformations that reduce the number of extrapolations to just two.

\section{Supersymmetry-preserving mass deformations}
\label{sec:susy-def}

We introduce supersymmetry-preserving deformations \cite{Bonelli:2002mb}, which includes a Myers term, to the original IKKT model $(S_{\rm IKKT})$. We obtain the following deformed model 
\beq
\label{eqn:susy-deformed-ikkt-action}
S = S_{\rm IKKT} + S_{\Omega},~~{\rm with}~~S_{\rm  \Omega} = N~ {\rm tr} \left( M^{\mu \nu} X_{\mu} X_{\nu} +i N^{\mu \nu \sigma} X_\mu \left[X_{\nu}, X_{\sigma}  \right]  +\frac{i}{8} \psib N_3 \psi \right),
\eeq
where $N_3 = \Gamma^{\mu \nu \sigma} N_{\mu \nu \sigma}$, with $N^{\mu \nu \sigma}$ denoting a totally anti-symmetric tensor, and $M_{\mu \nu}$ is the mass matrix. The deformed model preserves half of the supersymmetry and is invariant under the following transformation
\bea
\label{eqn:susy-deformed-ikkt-transformations}
\delta X^{\mu} = -\hf \overline{\varepsilon} \Gamma^{\mu} \psi,~~ \delta \psi = \qtr \left[ X^{\mu}, X^{\nu} \right] \Gamma_{\mu \nu} \varepsilon - \frac{i}{16} X^{\mu} \left( \Gamma_{\mu} N_3 + 2N_3 \Gamma_{\mu} \right)\varepsilon, 
\eea 
provided a mass/flux constraint, 
$\left[ N_3(\Gamma^{\mu} N_3 +2 N_3 \Gamma^{\mu}) - 4^3 M^{\mu \nu} \Gamma_{\nu} \right]\varepsilon = 0$, is satisfied. A straightforward solution to this constraint is to consider 
\beq 
\label{eqn:susy-deformed-ikkt-solution}
N_{3} = -\Omega\Gamma^{8}{\Gamma^{9}}^{\dagger}\Gamma^{10}, ~N^{\mu \nu \sigma} =  \frac{\Omega}{3!} \sum_{\mu,\nu,\sigma=8}^{10}{\epsilon^{\mu \nu \sigma}}  ~~{\rm and} ~~ M =  \frac{\Omega^2}{4^3} \left( \mathbb{I}_7 \oplus 3\mathbb{I}_3  \right), 
\eeq
which explicitly breaks the ten-dimensional rotational symmetry SO(10) to SO(7) $\times$ SO(3). One can obtain the original IKKT matrix model and study the spontaneous breaking of rotational symmetry by extrapolating $\Omega \to 0$ in the large-$N$ limit. 

The fermion bilinear deformation modifies the fermion operator in the following manner:
\bea
\mathcal{M}_{\alpha a \beta b}  \to \tilde{\mathcal{M}}_{\alpha a \beta b} &=& \frac{N}{2}  \Gamma_{\alpha \beta}^\mu{\rm tr} \left( X_\mu \left[t^a,t^b \right] \right) -\frac{i\Omega N}{8} \left( \Gamma^{8} {\Gamma^{9} }^{\dagger}\Gamma^{10} \right)_{\alpha\beta} \delta_{ab}.   
\eea
\begin{figure}[htbp]
	\centering
	\includegraphics[width=.32\textwidth,origin=c,angle=0]{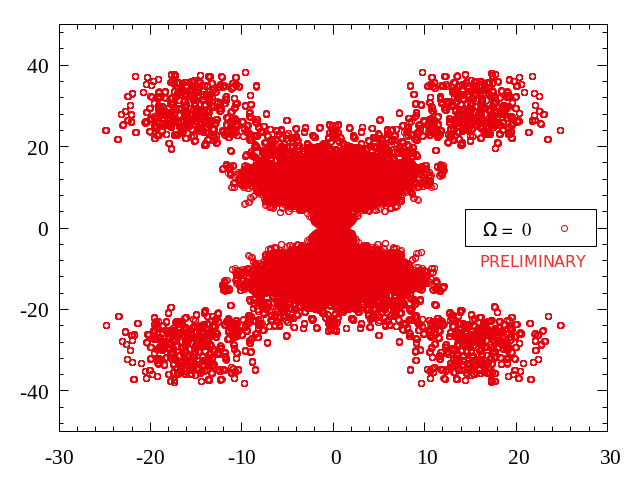}
	\includegraphics[width=.32\textwidth,origin=c,angle=0]{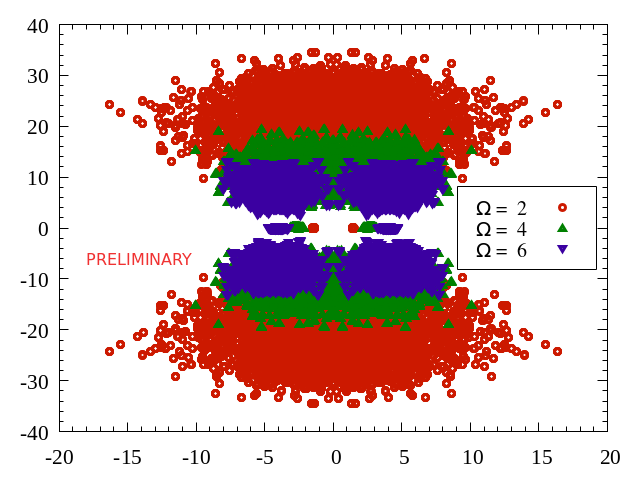}
	\includegraphics[width=.32\textwidth,origin=c,angle=0]{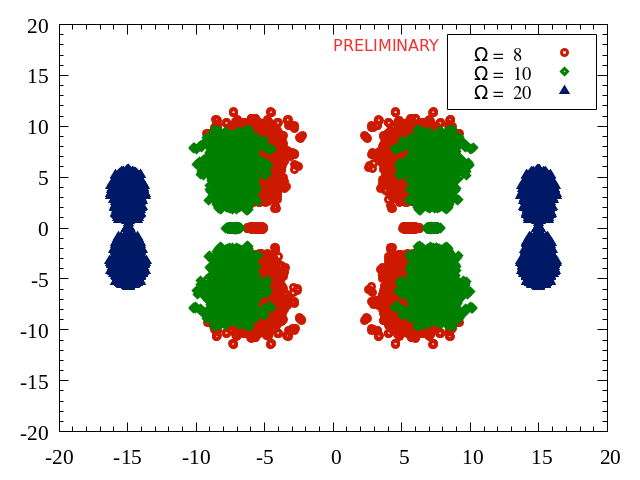}
	\caption{IKKT model with SUSY-preserving mass deformations. Scatter plot of real versus imaginary part of the eigenvalues of the fermion operator $\mathcal{M}$. The plots are for various mass deformation parameters $\Omega$ and fixed $N = 6$.} 
	\label{fig-eigen-nc4.png}
\end{figure}
In Fig. \ref{fig-eigen-nc4.png}, we plot the eigenvalue distribution of the fermion operator $\mathcal{M}$ for the SUSY-preserving mass deformed IKKT model. The singular-drift problem is apparent for mass deformation parameter $\Omega = 0$, that is, the original IKKT model. As we increase $\Omega$, the trend suggests that the eigenvalue distribution shifts further away from the origin. These results strongly indicate that SUSY-preserving mass deformations evade the singular-drift problem. 

\subsection{Bosonic IKKT deformed model with Myers term}
\label{subsec:bos-ikkt-myers}

We append the bosonic Gaussian mass deformation terms and a Myers term to the bosonic IKKT matrix model. The action of the deformed model reads $S_{\rm b} = S_{\rm bIKKT} + S_{\rm G} +S_{\rm Myers}$, where 
\bea
S_{\rm G} = \frac{\Omega^2 N}{4^3}~ {\rm tr} \left( \sum_{i=1}^{7} X_{i}^2 + 3 \sum_{a=8}^{10}  X_{a}^2 \right) {\rm ~and~} S_{\rm Myers} = \frac{i\Omega N}{3!}~ {\rm tr} \left( \sum_{a,b,c=8}^{10} X_{a} \left[X_{b}, X_{c}  \right] \right).
\eea
We perform complex Langevin simulations for various mass deformation parameters $\Omega$ and investigate whether the ten-dimensional rotational symmetry is intact in the $\Omega \to 0$ limit. We notice that the order parameter $\lambda_\mu(\Omega) $ has an inverse order dependence on mass deformation parameter $\Omega$. As a consequence, $\lambda_\mu (\Omega)$ blows up in the limit $\Omega \to 0$. To resolve this issue, we consider the normalized extent values defined as
\beq
\langle \rho_{\mu}(\Omega) \rangle \equiv \Big \langle  \frac{\lambda_\mu (\Omega) \rangle}{\sum_{\mu}\lambda_\mu (\Omega) } \Big \rangle.
\eeq
The normalized extents cancel a significant part of the dependency on the deformation parameter. In the case of broken SO(10) symmetry, the normalized extents $\rho_\mu$ will not be equal in all directions. 
\begin{figure}[htbp]
	\centering
	\includegraphics[width=.4\textwidth,origin=c,angle=0]{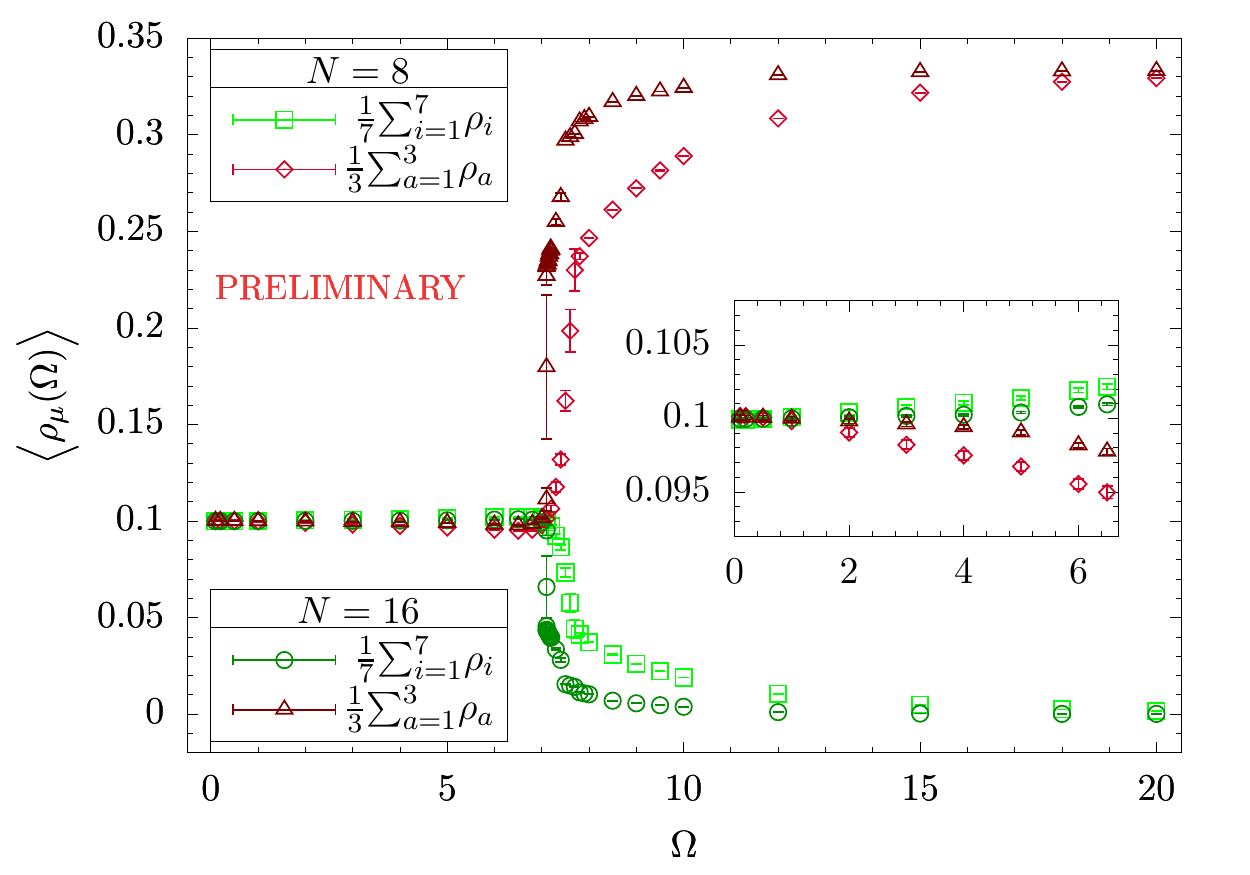}
	\includegraphics[width=.4\textwidth,origin=c,angle=0]{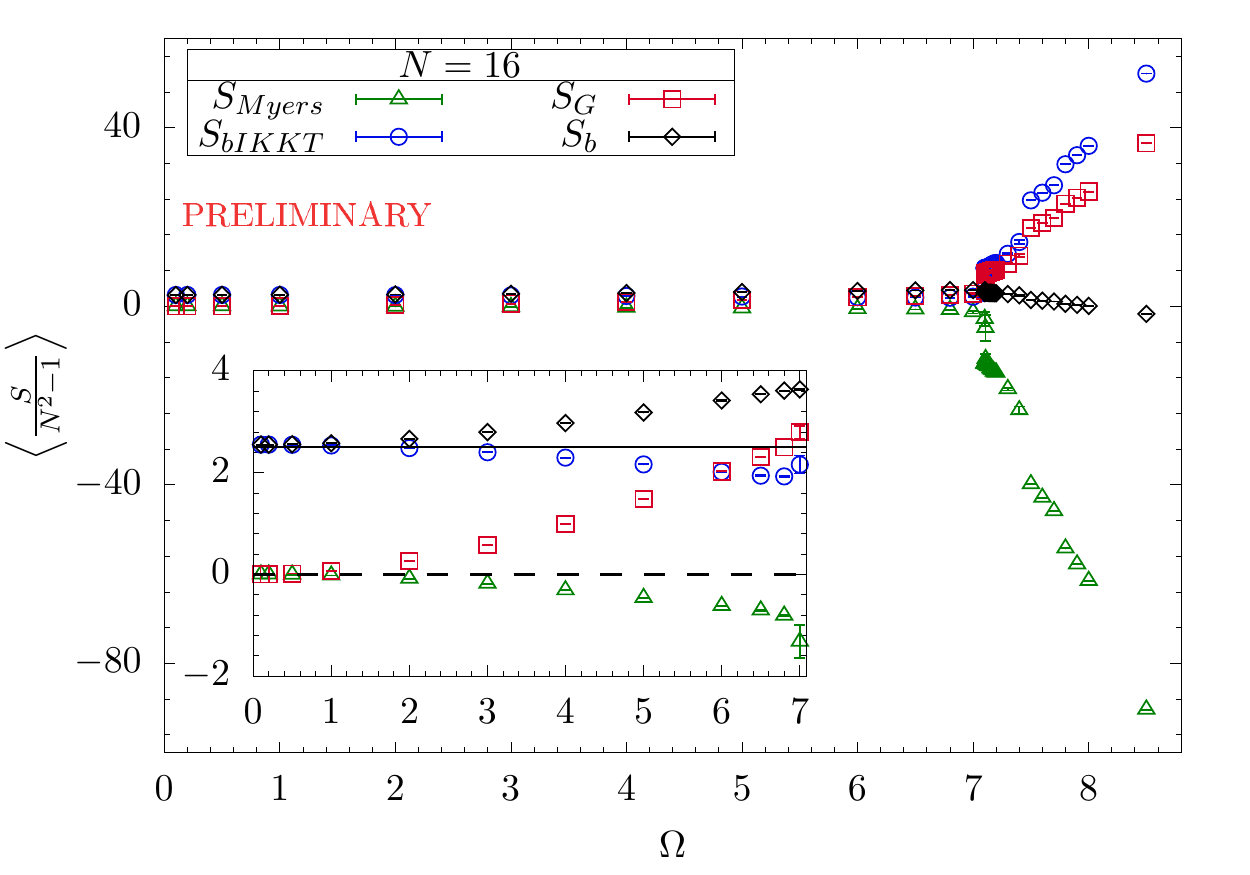}
	\caption{Deformed bosonic IKKT model with a Myers term. (Left) The averaged extents, $\frac{1}{7}\sum_{i=1}^{7} \rho_{i}(\Omega)$ and $\frac{1}{3}\sum_{a=8}^{10} \rho_{a}(\Omega)$ versus mass-deformation parameter $\Omega$ for $N=8, 16$. (Right) The bosonic action terms versus mass-deformation parameter $\Omega$ for $N = 16$.} 
	\label{fig:bos-myers}
\end{figure}

In this model, we observe an explicit symmetry breaking of SO(10) $\to$ SO(7) $\times$ SO(3) for large enough $\Omega$ values and thus, we have considered the averaged extents, that is, $\frac{1}{7}\sum_{i=1}^{7} \rho_{i}(\Omega)$ and $\frac{1}{3}\sum_{a=8}^{10} \rho_{a}(\Omega)$ as the order parameters. The averaged extents are shown on the left panel of Fig. \ref{fig:bos-myers}. In the limit $\Omega \to 0$, the two averaged extents converge, and the SO(10) symmetry of the original bosonic IKKT model is restored. These results demonstrate that the bosonic mass deformation and the Myers term do not play any role in the SSB of SO(10) symmetry. We also notice a first-order phase transition around $\Omega \sim 7.1$ for $N = 16$. We believe this is a consequence of the change in the saddle point configurations due to the Myers term. On the right panel of Fig. \ref{fig:bos-myers}, we see that the dominant nature of the Myers term is apparent after $\Omega \sim 7.1$. The inset plot shows that in the limit $\Omega \to 0$, the contributions from the Gaussian deformation and the Myers term vanish, and we obtain the bosonic IKKT model.  

\subsection{IKKT model with supersymmetry-preserving mass deformations}
\label{subsec:ikkt-susy-def}

\begin{figure}[htbp]
	\centering
	\includegraphics[width=.49\textwidth,origin=c,angle=0]{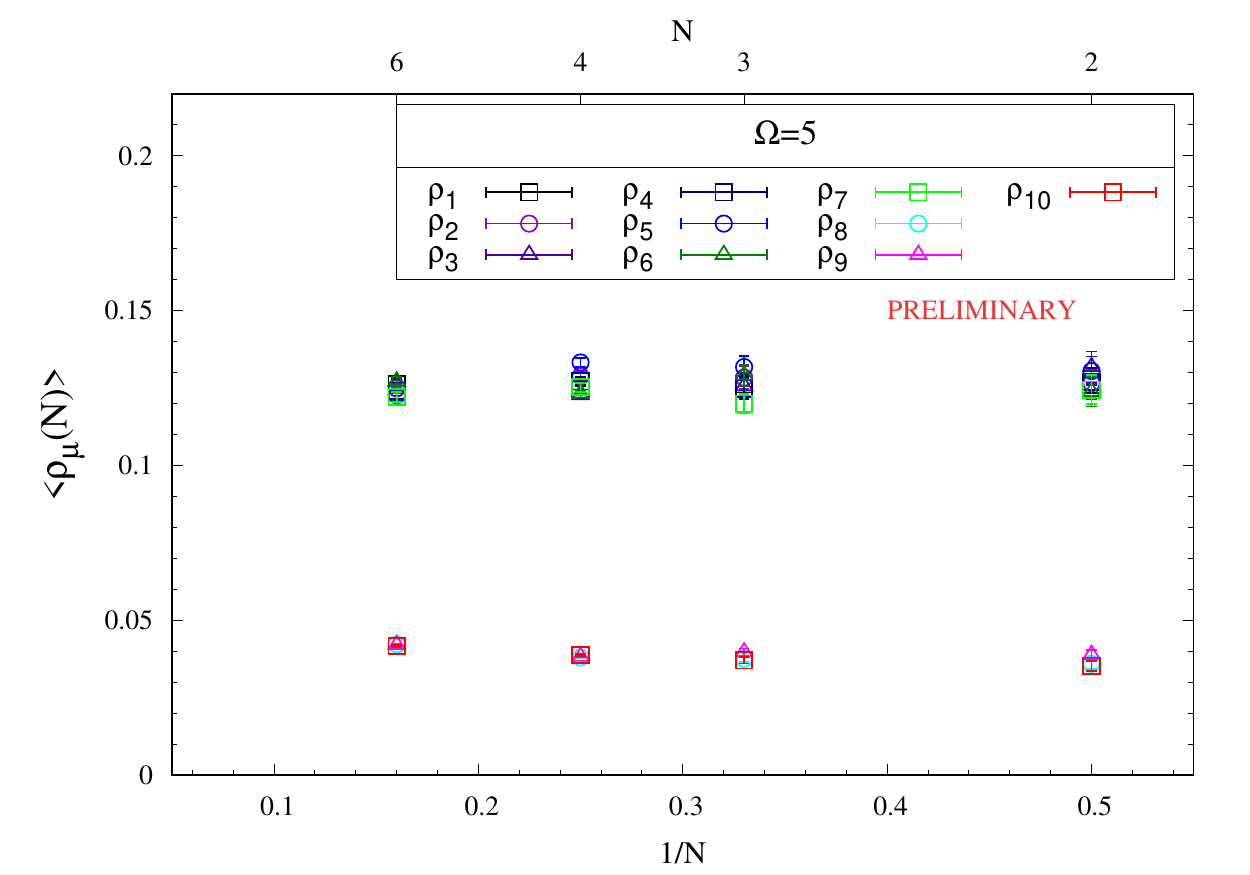}
	\includegraphics[width=.49\textwidth,origin=c,angle=0]{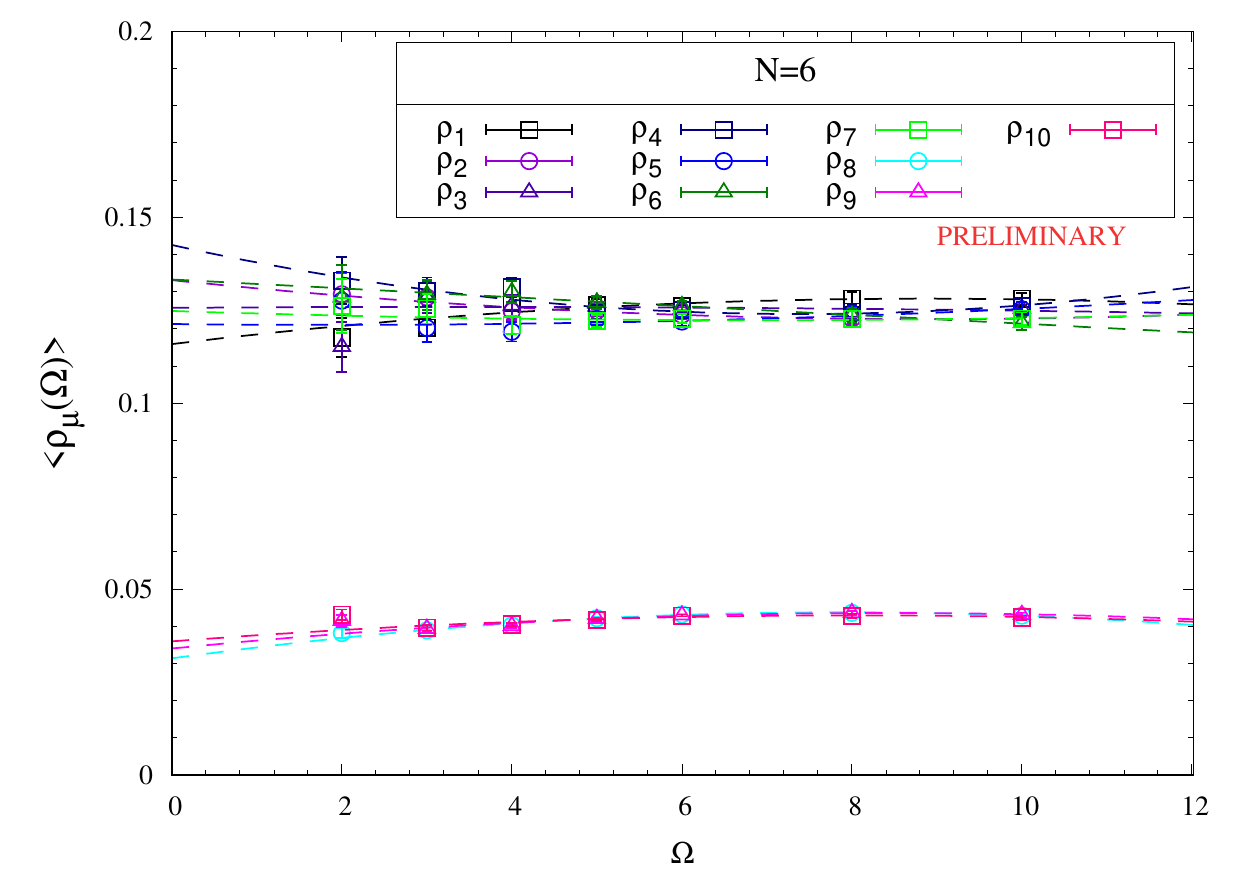}
	\caption{IKKT model with SUSY-preserving mass deformations. (Left) The normalized extents (order parameter) $\rho_{\mu}$ versus $N$ for fixed $\Omega = 5$. (Right) The normalized extents $\rho_{\mu}$ versus $\Omega$ for fixed $N = 6$. } 
	\label{fig:rho-N-Omega}
\end{figure}

This section reports our preliminary results from the complex Langevin simulations of the IKKT model with SUSY-preserving mass deformations. On the left panel of Fig. \ref{fig:rho-N-Omega}, we plot the normalized extents $\rho_\mu$ for fixed $\Omega = 5$ and various matrices of size $N$. For a large enough $\Omega$ value, we observe an explicit SO(7) $\times$ SO(3) symmetry breaking. Our finite-$N$ results suggest that the extents $\rho_\mu$ are almost independent of $N$, but we require large-$N$ computations to comment on the exact behavior concretely. 

The estimation of $\mathcal{M}^{-1}$ has a computational time complexity of $O(N^6)$ and is the bottleneck of the algorithm. In this preliminary study, we consider $N = 6$ as the large-$N$ limit and take the mass deformation parameter $\Omega \to 0$ limit on the right panel of Fig. \ref{fig:rho-N-Omega}. The complex Langevin simulations become unreliable for $\Omega < 2$. In limit $\Omega \to 0$, we recover the original IKKT matrix model, and even for $N = 6$, the spontaneous breaking of SO(10) $\to$ SO(7) $\times$ SO(3) is apparent. Interestingly, we notice that the SO(7) symmetry appears to further break down into smaller subgroups as $\Omega \to 0$, indicating a SO$(d)$ symmetric vacuum with $d < 7$. To investigate the exact nature of the symmetric vacuum of the IKKT matrix model, we need to consider large-$N$ extrapolations. 

\section{Conclusions and future prospects }
\label{sec:conclusion}

In this work, using the complex Langevin method, we have performed a first-principles study of the Euclidean IKKT matrix model. The main objective was to inspect the spontaneous symmetry breaking of ten-dimensional rotational symmetry. For the bosonic IKKT model, we did not see any signs of SSB, which is consistent with the previous Monte Carlo and $1/D$ expansion studies. In our simulations, we encountered the singular-drift problem. The preliminary results reported in this work suggest that adding supersymmetry-preserving mass deformations can successfully evade this singular-drift problem. We have also investigated the bosonic IKKT deformed model with the Myers term and found that the Gaussian deformation and Myers terms do not play any role in the SSB of SO(10) symmetry.

For the IKKT matrix model with SUSY-preserving deformations, our analysis indicates that the phase of the Pfaffian does indeed trigger the SSB of ten-dimensional rotational symmetry. For $N = 6$, we saw that the SO(7) symmetric vacuum was realized. We have also observed hints toward smaller subgroups SO$(d)$ symmetric vacua with $d < 7$. We plan to carry out a more robust large-$N$ analysis to find the exact nature of the vacuum. We are considering efficient techniques to compute the $\mathcal{M}^{-1}$ operator. Stochastic estimation of the fermion gradient is one such alternative. We hope to report the results of ongoing simulations soon.

{\bf Acknowledgements:} 
AK was partially supported by IISER Mohali and the Council of Scientific and Industrial Research (CSIR), Government of India, Research Fellowship (No. 09/947(0112)/2019-EMR-I). The work of AJ was supported in part by the Start-up Research Grant (No. SRG/2019/002035) from the Science and Engineering Research Board (SERB), Government of India, and in part by a Seed Grant from the Indian Institute of Science Education and Research (IISER) Mohali. PK was partially supported by the INSPIRE Scholarship for Higher Education by the Department of Science and Technology, Government of India. We acknowledge the National Supercomputing Mission (NSM) for providing computing resources through the PARAM Smriti supercomputing system at NABI Mohali.

\bibliography{pos_ikkt_clm.bib}
\bibliographystyle{JHEP.bst}

\end{document}